\documentclass{article}

\usepackage{emulateapj}
\usepackage{graphics}

\def\linebreak{\hfil\break}

\def\
{\hfil\linebreak}

\def\etal{{\it et al}. }
\def\degree{\ifmmode {^\circ}\else {$^\circ$}\fi}
\def\mum{\ifmmode {\rm \mu {\rm m}}\else $\rm \mu {\rm m}$\fi}
\def\arcsec{\ifmmode ^{\prime \prime}\else $^{\prime \prime}$\fi}

\def\inch{\ifmmode ^{\prime \prime}\else $^{\prime \prime}$\fi}
\def\arcmin{\ifmmode ^{\prime}\else $^{\prime}$\fi}

\def\msun{\ifmmode {\rm M_{\odot}}\else $\rm M_{\odot}$\fi}
\def\mearth{\ifmmode {\rm M_{\oplus}}\else $\rm M_{\oplus}$\fi}
\newbox\grsign \setbox\grsign=\hbox{$>$} \newdimen\grdimen \grdimen=\ht\grsign
\newbox\simlessbox \newbox\simgreatbox
\setbox\simgreatbox=\hbox{\raise.5ex\hbox{$>$}\llap
     {\lower.5ex\hbox{$\sim$}}}\ht1=\grdimen\dp1=0pt
\setbox\simlessbox=\hbox{\raise.5ex\hbox{$<$}\llap
     {\lower.5ex\hbox{$\sim$}}}\ht2=\grdimen\dp2=0pt

\begin{document}

\submitted{The Astrophysical Journal Letters, submitted}

\title{The Kuiper Belt and Olbers Paradox}

%
\author{Scott J. Kenyon}
\affil{Smithsonian Astrophysical Observatory, 60 Garden Street, 
Cambridge, MA 02138}
\affil{skenyon@cfa.harvard.edu}
\author{and}
\author{Rogier A. Windhorst}
\affil{Dept of Physics \& Astronomy, Arizona State University, Box 871504, Tempe, AZ 85287-1504}
\affil{Rogier.Windhorst@asu.edu}
\submitted{Re-submitted to the Astrophysical Journal Letters, 5 October 2000}
%


\begin{abstract}

We investigate the constraints that Olbers Paradox, applied to the 
Zodiacal Background as measured from space, sets on outer solar system 
objects. If extended to very faint limits, $R \sim$ 40--50 mag, the
steep optical number counts of Kuiper Belt objects (KBOs) at 
$R \lesssim$ 26 imply an infinitely bright night sky. Small KBOs with 
radii of $r \sim$ 1 $\mu$m to $r \sim$ 1 km must have a size distribution 
$n(r) \propto$ $r^{-a}$, with $a \sim$ 3.4 or smaller to satisfy the 
known limits on the sky-surface brightness at optical and far-infrared 
wavelengths.  Improved limits on the measured KBO surface brightness
can yield direct estimates of the albedo, temperature, and size 
distribution for small KBOs in the outer solar system.

\end{abstract}

\subjectheadings{Kuiper Belt -- solar system: formation}

\section{INTRODUCTION}

The darkness of the night sky -- Olbers Paradox -- is one of
astronomy's great mysteries.  The conclusion that an infinite 
universe full of stars and galaxies produces an infinitely 
bright sky is over four hundred years old.  The resolution of 
the Paradox, an expanding and infinite universe with a finite 
age, is now a central tenet of Big Bang Cosmology, where galaxy 
counts and the limiting sky background constrain the evolution 
of galaxies and the curvature of the universe (e.g., Driver \etal 1995, 
1996, 1998; Odewahn \etal 1996).  Here we show that local 
radiation sources, including the Kuiper Belt in our solar system, 
can also appear to violate Olbers Paradox.  Current limits on the 
sky background set interesting constraints on the mass and size 
distributions of Kuiper Belt objects (KBOs).

Recent observations have detected many small icy bodies in our
solar system outside the orbit of Neptune (\cite{luu98}).  
These KBOs consist of three dynamical classes:
(1) {\it classical} KBOs with roughly circular orbits and semimajor
axes of 41--47 AU,
(2) {\it resonant} KBOs, also known as Plutinos, in orbital resonance
with Neptune (Jewitt, Luu, \& Chen 1996), and
(3) {\it scattered} KBOs with eccentric orbits, $e \sim $ 0.5,  and 
perihelion distances of 30--38 AU (\cite{luu97}).
For an adopted optical albedo, $\omega \approx$ 0.04, KBOs with 
red magnitudes, R $\approx$ 20--26 mag, have radii of 50--500 km 
(\cite{luu98}).  

All optical measurements of individual KBOs are consistent with a 
simple function for the number counts:

\begin{equation}
{\rm log} ~ N (R) = \alpha (R - R_0) ~ ,
\end{equation}

\noindent
where $N$ is the cumulative number of KBOs per square degree.\footnote{We
use number counts throughout.  The luminosity function, defined as the
number per unit volume per magnitude, is more common in extragalactic 
astronomy.  In the Kuiper Belt, the analog to the luminosity function
may vary with distance from the Sun.} Recent analyses indicate 
$\alpha$ = 0.5--0.75 and $R_0$ = 23.2--23.5 mag in the Kron-Cousins 
photometric system (\cite{gla98}; Jewitt, Luu, \& Trujillo 1998; 
\cite{chi99}).
These results imply a differential size distribution,
$n(r) \propto r^{-a}$, where $a$ = 3.5--4.5 (e.g., \cite{luu98}).  
The total mass in large KBOs with radii 50--500 km is $\sim$ 
0.1--0.2 \mearth~for the 
classical and resonant KBOs (Jewitt \etal 1998; Luu \& Jewitt 1998) 
and $\sim$ 0.05 \mearth~for the scattered KBOs (Trujillo \etal 2000).
Coagulation models which form large KBOs from mergers of much smaller 
objects naturally produce size distributions, $n(r) \propto r^{-4}$,
and total masses of large KBOs consistent with the R-band observations 
(\cite{kl99a},b).  

Fits to the optical counts of KBOs can conflict with measurements of the 
known, finite sky brightness. Windhorst, Mathis, \& Keel (1992)
and Windhorst \etal (1994, 1998) quote on-orbit surface brightness 
measurements with the Hubble Space Telescope (HST) Wide Field/Planetary 
Camera 1 (WF/PC-1) and Wide Field Planetary Camera 2 (WFPC2) at ecliptic 
latitude $\approx$ 70\degree~for V, $\mu_V \approx$ 23.2 mag arcsec$^{-2}$, 
and for I, $\mu_I \approx$ 22.2 mag arcsec$^{-2}$.  All optical surface 
brightnesses throughout have units of Vega mag arcsec$^{-2}$.
The surface brightness in the ecliptic plane is roughly 1 mag 
arcsec$^{-2}$ brighter, $\mu_V \approx$ 22.35 mag arcsec$^{-2}$ 
for ecliptic longitude $\lambda \approx$ 110\degree--140\degree~and 
ecliptic latitude $\beta$ = 0\degree~(Table 6.2 in the WFPC2 Handbook;
Biretta \etal 2000).  For KBOs with a Kron-Cousins color of the Sun, 
V--R = 0.37 mag (\cite{jw98a}; \cite{dav00}), the limit to the 
R-band surface brightness from KBOs in the ecliptic plane is 
$\mu_R = 22$ mag arcsec$^{-2}$.
Using equation (1), faint KBOs reach this limit at R $\approx$ 45--55 mag
(see below). Thus, the KBO number counts cannot follow equation (1) to 
arbitrarily faint magnitudes.

The far-infrared (far-IR) background radiation place additional 
constraints on the KBO population.  Studies of the collisional 
evolution of KBOs (e.g., Backman, Dasgupta, \& Stencel 1995; 
\cite{tep99}) show that the size distribution of objects with radii 
of several microns to several millimeters should have a power law 
index $a \approx$ 3.5.  For small Kuiper Belt grains with temperatures 
$ T_{KBO} \sim$ 40 K and a total mass in small grains of $\sim$ 
$10^{-5}$ \mearth, far-IR fluxes predicted from these models 
typically fall below reliably available background measurements 
at wavelengths longer than $\sim$ 10 $\mu$m, 
$I_{\nu} ({\rm FIR}) \lesssim$ 1--2 $\times ~ 10^6$ Jy sr$^{-1}$ (\cite{fix98}; 
\cite{hau98}).  For grains with $T_{KBO} \lesssim$ 100 K, we show 
below that this limit provides a better constraint on the properties 
of small KBO grains than current surface brightness limits set 
by {\it HST}.  However, with 1$\sigma$ sensitivity limits of 
$\mu_V \approx$ 27 mag arcsec$^{-2}$ and $\mu_I \approx $ 
25.4 mag arcsec$^{-2}$, WFPC2 images can, in principle, detect 
a Kuiper Belt small grain population which emits less light than 
the currently available far-IR limits.

Our goal is to quantify the relationship between the optical number 
counts for bright KBOs and the measured sky-surface brightness at 
optical and far-IR wavelengths.  In \S2, we use equation (1) to relate 
the observed sky-surface brightness to the slope of the optical counts.  
In \S3, we derive a physical model relating observables to the KBO
size distribution. We conclude in \S4.

\section{THE OPTICAL SURFACE BRIGHTNESS OF KBOs}

For equation (1), the R-band surface brightness for KBOs with magnitude 
brighter than R is:

\begin{equation}
\mu_R = 41.03 - 2.5 ~ {\rm log} \left ( \frac{\alpha}{\alpha - 0.4} \right ) + (1 - 2.5 \alpha) ~ (R - R_0)
\end{equation}

\noindent
for $\alpha > 0.4$.  The surface brightness at other optical and 
near-IR bands follows from assumptions for $\omega$ and the typical 
color of a KBO.  To relate limits on thermal emission to the optical 
sky brightness, we assume that $F_{tot}$ is the total flux density 
in erg cm$^{-2}$ s$^{-1}$ sr$^{-1}$ received at Earth from KBOs.  
If the albedo for KBOs is independent of size, then the R-band surface 
brightness in mag arcsec$^{-2}$ is: 

\begin{equation}
\mu_R = -2.5 ~ {\rm log} \left ( \frac{\omega F_{tot}}{A_0 F_0} \right ) - BC(R), 
\end{equation}

\noindent
where $A_0$ is the number of arcsec$^2$ in a steradian, 
$F_0$ = $2.48 \times 10^{-5}$ erg cm$^{-2}$ s$^{-1}$ is the 
zero point (Allen 1976), and BC(R) = +0.17 is the bolometric 
correction (Kenyon \& Hartmann 1995).  
The thermal radiation received from KBOs is $ (1 - \omega) F_{tot}$. 
If KBOs radiate as blackbodies, an absolute upper limit to 
the flux at frequency $\nu$ is the flux at the Planck peak,
$I_{\nu,m} \approx 10^{-11}$ $T_{KBO}^{-1} (1 - \omega) F_{tot}$, 
where $T_{KBO}$ is measured in Kelvins.  
If we solve equation (3) for $F_{tot}$, the thermal emission from KBOs 
depends only on the albedo, the surface brightness, and the temperature:

\begin{equation}
I_{\nu} ({\rm FIR}) = 9.5 \times 10^{17 - 0.4\mu_R} ~ T_{KBO}^{-1} ~ \left ( \frac{1 - \omega}{\omega} \right )~ {\rm Jy ~ sr^{-1}} ~ .
\end{equation}

\noindent
For a fixed optical brightness, equation (4) predicts that 
hot KBOs produce less thermal emission than cold KBOs.  
This counter-intuitive result is correct, because hot KBOs 
must have a smaller surface area to emit the same bolometric 
flux as cold KBOs.

Equation (4) has three main uncertainties.  From the apparent spread
in the optical colors of KBOs (\cite{jw98a}; \cite{dav00}), the bolometric 
correction is uncertain by $\pm$ 0.2 mag.  Large KBOs and comets with 
$\omega \approx$ 0.04 (e.g., \cite{luu98}) contribute little to the 
optical surface brightness or the thermal emission. If $\omega \sim$ 
0.2--0.5 for the smallest KBO grains (Backman \& Paresce 1993; 
\cite{bac95}), $I_{\nu}$(FIR) is uncertain by a factor of 4.  
Small KBO grains with $r \lesssim$ 100 \mum~are inefficient radiators 
and are hotter than larger KBOs at the same distance from the Sun 
(see Backman \etal 1995; Teplitz \etal 1999).  For constant $F_{tot}$, 
$I_{\nu}$(FIR) for small grains is a factor of 2--5 less than 
the limit in equation (4). We conclude that equation (4) with 
$\omega$ = 0.5 and $T_{KBO}$ = 40 K is a reasonable estimate for thermal 
emission from KBOs with optical number counts defined by equation (1).

The solid lines in Figure 1 indicate how the optical and far-IR surface
brightness increase with fainter KBO R-band magnitude, assuming 
$R_0$ = 23.25 mag and $\alpha$ = 0.5, 0.6, and 0.7. 
The dashed lines in each panel show observed limits, 
$I_{\nu}({\rm FIR})$ = $10^6$ Jy sr$^{-1}$ and
$\mu_R \approx$ 22 mag arcsec$^{-2}$.  These simple models,
which fit the optical counts for R $\le$ 26--27 mag, yield optical 
and far-IR backgrounds much larger than current limits.  
The vertical tic marks in Figure 1 indicate approximate radii
corresponding to selected R magnitudes for KBOs at 40 AU.  The limits
on optical and far-IR surface brightness require that the size
distribution of KBOs turns over for objects with $r \lesssim$ 10 m.

To allow for a shallower size distribution of small KBOs, we consider
the surface brightness distribution for number counts that follow a 
broken power law:

\begin{equation} 
{\rm log} ~ N (R) = \begin{array}{l l}
	\alpha_1 (R - R_0) & R < R_1 \\
                           & \\
	\alpha_1 (R_1 - R_0) + \alpha_2 (R - R_1) & R \ge R_1 \\
\end{array}
\end{equation}

\noindent
Similar forms for broken power laws have been derived for other solar 
system bodies from observations (e.g., \cite{wei82}; \cite{bai88}) 
and theory (e.g., \cite{gre84}; \cite{wet93}).  Weissman \& Levison 
(1997) previously proposed a broken power law for KBOs to satisfy 
constraints from the optical counts and the supply of short period comets.

To construct surface brightness distributions for a broken power law, 
we adopt $\alpha_1$ = 0.6 and $R_0$ = 23.25 mag to match the optical number 
counts.  Observations currently require $R_1 \ge$ 26--27 mag to fit 
equation (1) to the optical data. We adopt $R_1$ = 30 mag for 
simplicity.  The dashed lines in Figure 1 show how this model compares 
with the observed limits for $\alpha_2$ = 0.4--0.5.  Models with 
$\alpha_2 \lesssim$ 0.48 satisfy the optical and far-IR 
sky-background constraints.  The allowed power law slope becomes 
flatter as the knee in the number counts moves to fainter magnitudes.  

\section{PHYSICAL MODEL FOR SURFACE BRIGHTNESS}

To understand how the constraints on the optical counts relate to the
size distribution of KBOs, we construct a physical model for the 
observed surface brightness.  We adopt a broken power law size
distribution:

\begin{equation} 
n (r) = \begin{array}{l l}
	n_0 (r/r_0)^{-a_1} & r > r_0 \\
                           & \\
	n_0 (r/r_0)^{-a_2} & r \le r_0 \\
\end{array}
\end{equation}

\noindent
and assume objects are distributed in a ring around the Sun with a
surface density, $\Sigma \propto A^{-\beta}$. The ring has an inner
radius $A_1$ = 40 AU and an assumed outer radius $A_2$ = 50 AU.  We adjust 
$n_0$ and $a_1$ to match the observed optical counts, $\alpha$ = 0.6 
and $R_0$ = 23.25 mag, for an adopted $\omega$ = 0.04 and slope parameter
$g = 0.15$ in the standard two-parameter magnitude relation for
sunlight scattered by asteroids (\cite{bow89}).  The slope parameter
$g$ relates the brightness of an asteroid at any solar phase angle 
$\theta$ to the brightness at opposition, when $\theta$ = 0\degree.
We assume $\theta =$ 0\degree~for these calculations.  The 
predicted $\mu_R$ 
follows from a sum over all objects projected into a box with an 
area of 1 arcsec$^2$.  To predict thermal emission from these objects, 
we use equations (4) and (5) of Backman \& Paresce (1993) to derive 
grain temperatures as a function of distance from the Sun.
Large objects absorb and radiate as blackbodies with $T_{KBO}$ = 
278 $(1 - \omega)^{1/4} (A/{\rm 1 ~ AU})^{-1/2}$; small grains 
radiate with an emissivity $\epsilon \propto \lambda^{-1}$
and have $T_{KBO}$ = 468 
$(1 - \omega)^{1/4} (A/{\rm 1 ~ AU})^{-2/5} r^{-1/5}$ where $r$ is 
the radius in microns (see also \cite{tep99}).  The thermal emission 
$I_{\nu} ({\rm FIR)}$ follows from the appropriate sum over all bodies 
in a solid angle of 1 steradian.

This model ignores the complex orbital distribution of KBOs at 35--50 AU, 
where the bright end of the optical number counts is directly measured.  
Properties of KBOs beyond 50 AU are unknown.  The relationship between 
$\mu_R$ and $I_{\nu}$(FIR) is independent of our approximations and 
uncertainties.  Our results are insensitive 
to $\beta$, because we fit the observed optical counts.
The model predicts smaller far-IR fluxes for grains with an
emissivity $\epsilon \propto \lambda^{-2}$; current observations 
cannot distinguish between possible emissivity laws.  Although 
specific predictions for the optical and far-infrared surface brightness 
are sensitive to other adopted parameters, our main conclusions 
are independent of these assumptions except as indicated below.

The curves in Figure 2 show how the optical and 100 $\mu$m surface
brightness increase with fainter KBO R-band magnitude for models 
with $\beta$ = 1.5, $a_1 = 4$, and $a_2$ = 3.0, 3.25, and 3.5.  
Solid lines show results when all objects have $\omega$ = 0.04; 
dot-dashed lines show how the surface brightness changes when 
the albedo varies smoothly from $\omega$ = 0.04 for 
$r \ge$ 1 km to $\omega$ = 0.5 for $r \le$ 0.1 km.  As expected,
the larger albedo produces a brighter optical surface brightness, 
but a fainter far-IR surface brightness.  For models with $a_2$ = 3.5,
KBOs with a small constant albedo have a limiting optical surface 
brightness, $\sim$ 24.5 mag arcsec$^{-2}$, fainter than the observed 
sky brightness.  This limit, however, is detectable with {\it HST} 
if other background sources can be reliably subtracted.
If small KBOs have $a_2$ = 3.5 and a large albedo, the predicted 
$\mu_R$ exceeds the observed background at $R \sim$ 70 mag. This 
limit corresponds to objects with $r \sim$ 0.03 mm.  In both cases, 
the far-IR surface brightness exceeds the measured sky brightness 
for $\lambda \le$ 240 $\mu$m at $R \approx$ 65--70 mag.  The predicted
far-IR surface brightness lies below measured limits at longer wavelengths.

The far-IR surface brightness at 100 $\mu$m flattens in Figure 2 for 
R $\ge$ 70 mag, because grains with $r \lesssim$ 100 \mum~radiate
 inefficiently and are too hot to emit much 100 $\mu$m light.  If 
the albedo of these small grains remains large,
constraints on the KBO small grain population from optical data
are stronger than the COBE constraints for R $\lesssim$ 75 mag
(i.e., $r \le$ 1--3 $\mu$m). Uncertain optical properties for 
small KBO grains and rapid grain removal processes such as 
Poynting-Robertson drag make it difficult to use any observation 
to constrain the mass in KBO grains with $R \gtrsim$ 75 mag.

The results in Figure 2 show that the predicted surface brightness is 
very sensitive to $a_2$, the power law slope of the small object size
distribution.  Models with $a_2 \le$ 3.4 are consistent with the
surface brightness limits set by {\it COBE} for $\lambda \ge$ 10
$\mu$m and by {\it HST} at V and I.  Models with $a_2 \le$ 3.25 
predict background fluxes well below current detection limits at 
all wavelengths.

\section{DISCUSSION AND SUMMARY}

Our results demonstrate a clear relationship between the optical
and far-IR surface brightness from KBOs in the outer solar system.
The relation depends only on $\omega$ and $T_{KBO}$.  Direct 
measurements of $\mu_R$ and $I_{\nu}$ from KBOs should yield 
direct constraints on $\omega$ and $T_{KBO}$ for small grains.  
Observations of $\mu$ at other optical and near-IR wavelengths 
measure the colors of small KBOs.

The optical and far-IR surface brightness place useful constraints 
on the size distribution of small KBOs.  
To avoid violating Olbers Paradox at optical and far-IR wavelengths, 
the slope of the optical counts must change from $\alpha_1 \approx$ 
0.6 at $R \lesssim$ 27 mag to $\alpha_2 \lesssim$ 0.48 at 
$30 \le R \le 50$ mag.  With $a = 5 \alpha +1$, the slope of the KBO 
size distribution must change from $a_1 \approx$ 4 for $r \gtrsim$ 10 km 
to $a_2 \lesssim$ 3.4 for 1 m $\le r \le$ 1 km.  This conclusion is 
independent of assumptions about $\omega$, $T_{KBO}$, or the space 
distribution of KBOs.  If many small KBOs lie beyond 50 AU, the 
knee in the size distribution must occur at $ r \gtrsim$ 1 m.

Our upper limit for $\alpha_2$ is larger than the usual result 
for Olbers paradox, $\alpha_2 \le 0.4$.  In extragalactic astronomy,
galaxies fill an infinite volume nearly uniformly; galaxy counts 
with $\alpha \ge$ 0.4 always produce a divergent sky brightness.
For dust grains, the sky brightness converges when the largest 
objects produce all of the light ($a \le 3$, $\alpha \le$ 0.4);
the sky brightness diverges when the smallest objects produce all 
of the light ($a > 3$, $\alpha >$ 0.4).  Because dust grains and 
the disk of our solar system have finite sizes, the sky brightness
can converge for $\alpha > 0.4$.  For grains with $r \gtrsim$
0.01 $\mu$m at 40--50 AU, models with $\alpha_2 \le$ 0.47 do not 
violate Olbers Paradox. If Poynting-Robertson drag and radiation 
pressure are effective at removing grains with $r \le 1 ~ \mu$m, 
this limit is raised to $\alpha_2 \le$ 0.48.

These results support coagulation models for the formation of the
Kuiper Belt during the early evolution of our solar system 
(Kenyon \& Luu 1999a,b).  In this picture, 
small objects with $r \lesssim$ 1--100 m collide and merge to 
form larger objects.  In 10--100 million years, coagulation 
leads to a population of 1--1000 km objects with a size
distribution $n(r) \propto r^{-4}$.  Smaller objects have a shallower
size distribution, $n(r) \propto r^{-3.5}$ (Kenyon \& Luu 1999a;
\cite{bac95}). Orbital evolution of KBOs leads to large collisional 
velocities and gradual erosion of the Kuiper Belt (\cite{hol93}). 
Collisions between high velocity objects produce debris which adds 
to the `collisional tail' where $n(r) \propto r^{-3.5}$ (\cite{dav97}).
Poynting-Robertson drag and radiation pressure remove objects with 
radii less than 1 mm on timescales roughly proportional to the 
grain radius, $\sim 10^6$ yr ($r$/1 \mum) (\cite{bur79}; \cite{bac93}).
Grain removal may lead to a size distribution slightly flatter 
than $n(r) \propto r^{-3.5}$ for $r \sim$ 1 \mum~to 1 mm at 
35--50 AU\footnote{In the inner solar system at 1--2 AU, 
Poynting-Robertson drag and radiation pressure timescales 
are $\sim$ 1000 times shorter than in the Kuiper Belt,
which leads to rapid removal of grains with radii of 1 mm or
smaller.  Very small grains in the inner solar system thus make 
a negligible contribution to the brightness of the night sky.}.
Our analysis, together with those of previous investigations, 
demonstrates that the transition from a merger population with 
$n(r) \propto r^{-4}$ to a debris population with 
$n(r) \propto r^{-3.4}$ must occur at radii 10--1000 m, 
where collisions should  produce debris instead of mergers.

If other background sources can be reliably subtracted, current and 
future space missions can detect the integrated scattered light and 
thermal emission from small KBOs.  Direct detection of KBOs in 
the optical/near-IR or far-IR background would begin to provide 
stringent tests of collision models.  Constraints on the
variation of surface brightness with ecliptic latitude or
longitude would allow direct measurements of the scale height 
and orbital distributions of small KBOs.  In addition to possible
ground-based strategies, the sensitivity of deep HST WFPC2 images is 
now sufficient to improve constraints on the KBO optical background 
by a factor of ten. The {\it Space Infrared Telescope Facility} 
can improve the far-IR constraints by a similar factor. The 
{\it Next Generation Space Telescope} will provide direct 
measurements of individual KBOs near the proposed knee in the 
size distribution at $R \approx$ 28--31 mag and more accurate 
background measurements in the optical and near-IR.  These
and other facilities will yield better understanding of the
implications of Olbers Paradox in our solar system and beyond.

Comments from M. Geller, M. Holman, B. Marsden, and an anonymous
referee improved the accuracy and quality of the manuscript.

\vfill
\eject

\hskip 25ex
\epsffile{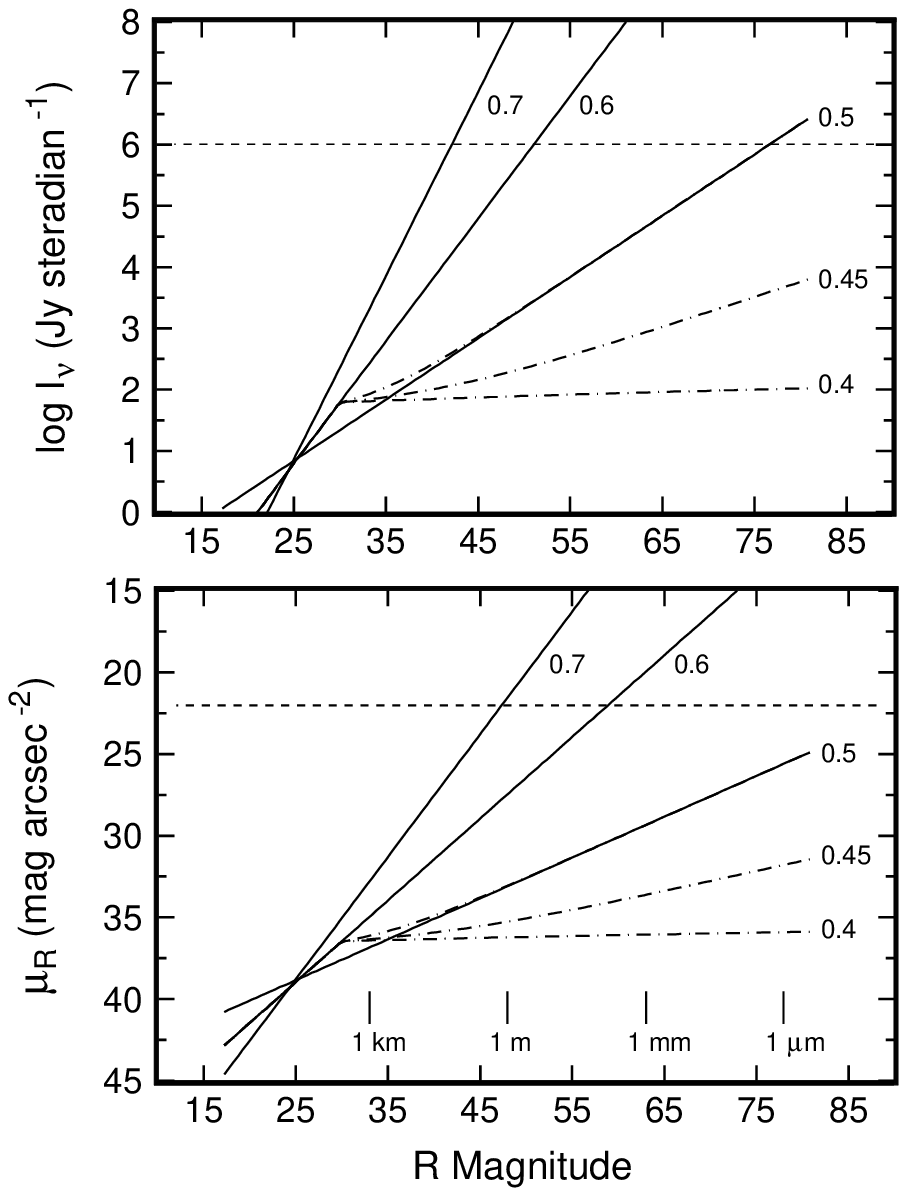}


\figcaption[Kenyon.fig1.eps]{Far-infrared and optical surface 
brightness as a function of R magnitude.  Each panel shows how
the integrated surface brightness increases with the R-magnitude 
limit for models with a single power law for the number counts 
(solid curves; equations 1, 2, and 4) or a broken power law fit
to the number counts (dot-dashed curves; equation 5).  Values 
for the power law slope, $\alpha$, appear to the right of each 
solid curve.  Dot-dashed curves have $\alpha_1$ = 0.6 and 
$\alpha_2$ as indicated. Each model is consistent with 
observations of the optical counts at $R \le$ 26--27.}

\vfill
\eject

\hskip 25ex
\epsffile{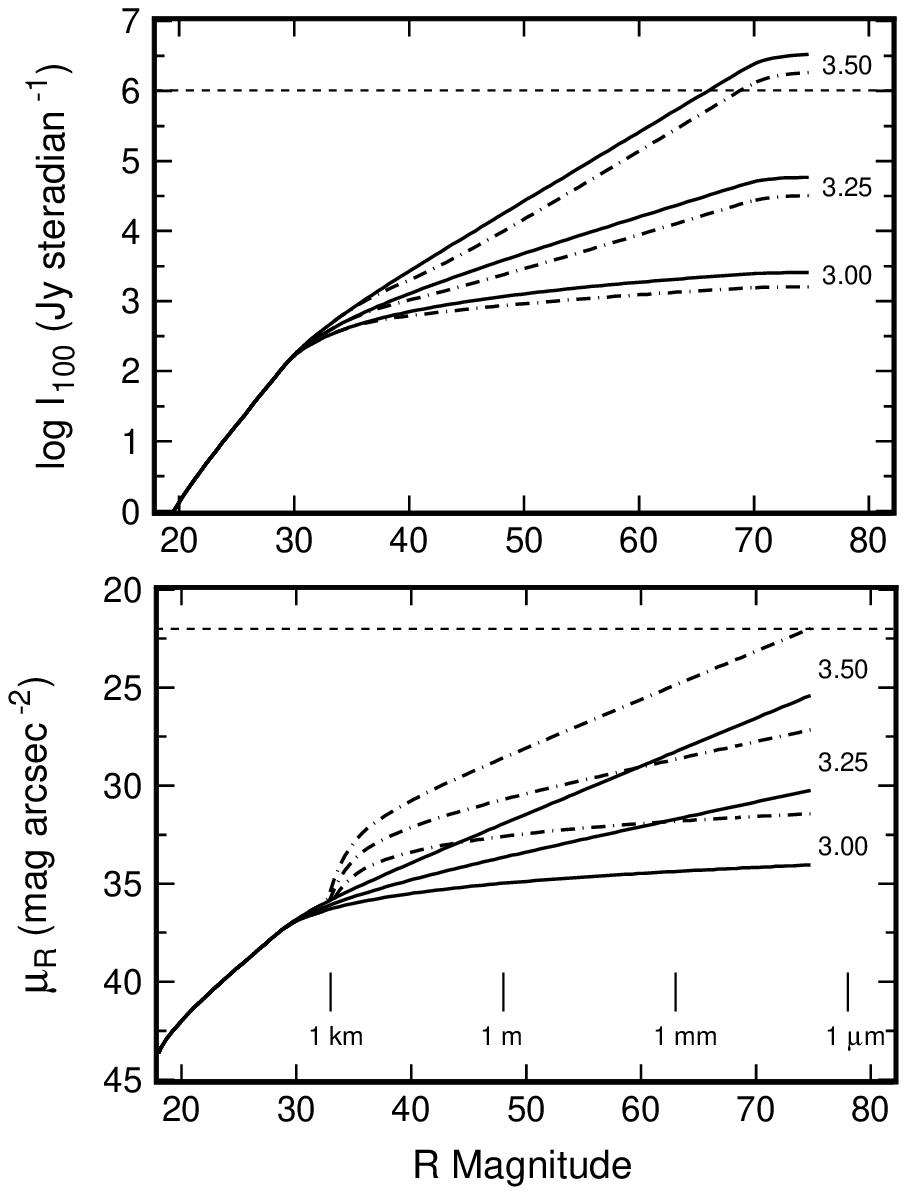}

\figcaption[Kenyon.fig2.eps]{As in Figure 1 for a physical KBO 
model.  The model assumes a broken power law size distribution,
equation (6), albedo $\omega$, and a surface density distribution 
for KBOs in a ring at 40--50 AU.  Solid curves show results for 
$a_1$ = 4, $\omega$ = 0.04, and $a_2$ as indicated at the right 
end of each curve.  Dot-dashed curves repeat this model for small
grains with larger $\omega$ as described in the text. Each model is
consistent with observations of the optical counts at $R \le$ 26--27.}

\end{document}